\documentclass[twocolumn,nofootinbib,superscriptaddress]{revtex4-1}

\usepackage[english]{babel}
\usepackage[OT1]{fontenc}
\usepackage[latin1]{inputenc}
\usepackage{xcolor}
\usepackage{latexsym,amsmath,amsfonts,amssymb,mathtools,mathrsfs}
\usepackage{footmisc} 
\usepackage{hyperref}

\newcommand\dfR{\boldsymbol{R}}

\newcommand\cofr{\boldsymbol{e}}
\newcommand\dimM{D}
\newcommand\mP{M_{\mathrm{P}}}
\newcommand\CosmC{\Lambda_0}
\newcommand\KMaxSym{\Lambda}

\begin{document}
\title{Inconsistencies in four-dimensional Einstein-Gauss-Bonnet gravity}

\author{Julio Arrechea}
\email{arrechea@iaa.es}
\affiliation{Instituto de Astrof\'isica de Andaluc\'ia (IAA-CSIC), Glorieta de la Astronom\'ia, Granada, Spain}

\author{Adri\`a Delhom}
\email{adria.delhom@uv.es}
\affiliation{Departamento de F\'{i}sica Te\'{o}rica and IFIC, Centro Mixto Universidad de Valencia - CSIC. Universidad de Valencia, Burjassot-46100, Valencia, Spain}

\author{Alejandro Jim{\'e}nez-Cano}
\email{alejandrojc@ugr.es}
\affiliation{Departamento de F\'{i}sica Te\'{o}rica y del Cosmos and CAFPE, Universidad de Granada, 18071 Granada, Spain}

\date{\today}
\begin{abstract}
We attempt to clarify several aspects concerning the recently presented four-dimensional Einstein-Gauss-Bonnet gravity. We argue that the limiting procedure outlined in [Phys. Rev. Lett. 124, 081301 (2020)] generally  involves ill-defined terms in the four dimensional field equations. Potential ways to circumvent this issue are discussed, alongside some remarks regarding specific solutions of the theory. We prove that, although linear perturbations are well behaved around maximally symmetric backgrounds, the equations for second-order perturbations are ill-defined even around a Minkowskian background. Additionally, we perform a detailed analysis of the spherically symmetric solutions, and find that the central curvature singularity can be reached within a finite proper time.
\end{abstract}

\maketitle

\section{Introduction}
It has been recently claimed \cite{Glavan2020} that there exists a theory of gravitation in four spacetime dimensions which fulfills all the assumptions of the Lovelock theorem \cite{Lovelock1972} yet not its conclusions. This was accomplished by formulating the Einstein-Gauss-Bonnet (EGB) theory in an arbitrary dimension $\dimM$ with a coupling constant for the Gauss-Bonnet term re-scaled by a factor of $1/(\dimM-4)$, as defined by the following action:
\begin{equation}\label{action}
    S=\int \mathrm{d}^{\dimM}x \sqrt{|g|}\left[-\CosmC + \frac{\mP^2}{2}R+\frac{\alpha}{\dimM-4}\mathcal{G}\right].
\end{equation}
Here $\Lambda_{0}$ is a cosmological constant term, $R$ is the Ricci scalar and $\mathcal{G}$ the Gauss-Bonnet  (GB) term. As it is well known, the GB term is a topological invariant only in $\dimM=4$ and not in higher dimensions, and it thereby generally yielding a non-trivial contribution to the field equations in arbitrary $\dimM$. In \cite{Glavan2020} it is claimed that the contribution of the Gauss-Bonnet (GB) term to the equations of motion is always proportional to a factor of $(\dimM-4)$, which in principle, compensates the divergence introduced in the coupling constant, thus allowing for a well defined $\dimM\to4$ \textit{limit} at the level of the field equations. It was argued that a non-trivial correction to General Relativity due to the GB term in \eqref{action} remains even in $\dimM=4$.

Since the above action is one of the celebrated Lovelock actions in arbitrary $\dimM$, it is stated in \cite{Glavan2020} that all the assumptions of Lovelock theorem hold, although the resulting field equations do violate the conclusions of the Lovelock theorem. This is accomplished by defining a 4-dimensional diffeomorphism-invariant theory satisfying the metricity condition and having second-order field equations that differ from those of General Relativity (GR). The authors of \cite{Glavan2020} then proceed to show the consequences of these modifications to GR in scenarios with a high degree of symmetry. Strong interest in this theory, which we will refer to as $\dimM\to4$ Einstein-Gauss-Bonnet (D4EGB, for short), has emerged recently. In particular, there has been an ongoing discussion \cite{Lu2020,Bonifacio2020,Tian2020,Hennigar2020,Kobayashi2020,Tian2020,Mahapatra2020,Fernandes2020,Ai2020,Shu2020} regarding the nature and/or the well-definiteness of D4EGB. In this article we also elaborate in this direction.

We now highlight some subtleties in the definition of D4EGB and in some of the solutions to its field equations. It was claimed in \cite{Glavan2020} that the contribution of the GB term (and not just its trace) to the field equations is proportional to $(\dimM-4)$, and that this implies the GB contribution to the field equations vanishes in four spacetime dimensions. The authors of \cite{Glavan2020} then consider a coupling constant with a factor of $1/(\dimM-4)$ that would \textit{regularise} the otherwise vanishing GB contribution, thereby yielding a finite correction to the four dimensional field equations. We show, in agreement with \cite{Gurses2020}, that in addition to a term proportional to $(\dimM-4)$, the GB term contributes to the field equations via an additional part from which no power of $(\dimM-4)$ can be factorized but which nonetheless vanishes identically in $\dimM=4$. 

Regarding tensor perturbations in D4EGB we reproduce the results of \cite{Glavan2020} for linear perturbations around a maximally symmetric background. This allows to determine that the theory only propagates a massless graviton and that the corrections to GR provided by the \textit{regularized} GB term only enter through a global $\alpha-$dependent factor multiplying the linear perturbation equations in GR. Nonetheless, we see that the field equations describing second-order perturbations contain ill-defined terms proportional to $1/(\dimM-4)$ that evaluate to an undetermined $0/0$ in the $\dimM\to4$ \textit{limit}, even around a Minkowskian background. Indeed, we argue that unless one is considering solutions with enough symmetry so as to force a specific combination of Weyl tensors to vanish in arbitrary dimensions a priori, the term that is not proportional to $(\dimM-4)$ renders the unperturbed D4EGB field equations ill-defined.

Finally, we comment on the geometries presented in \cite{Glavan2020} as the $\dimM\to4$ \textit{limit} of the spherically symmetric solutions for EGB theory in $\dimM\geq5$ found in \cite{Boulware1985}. We see that the claim made in \cite{Glavan2020} that no particle can reach the central curvature singularity in a finite proper time within these geometries does not hold for freely-falling particles with vanishing angular momentum. Furthermore, we will show that the \textit{regularized} D4EGB field equations are not well defined in spherically symmetric spacetimes unless the contribution that is not proportional to $(\dimM-4)$ is artificially stripped away from the field equations. Moreover, in the case in which this term is removed, we argue that the spherically symmetric geometries presented in \cite{Glavan2020} are not solutions of the remaining field equations in $\dimM=4$.

\section{The $\dimM\to4$ procedure}\label{sec:Dto4}

Let us first comment on whether the $\dimM\to4$ limit taken in \cite{Glavan2020} corresponds to a well-defined continuous process. To that end, consider the $k$-th order Lovelock term in an arbitrary dimension $\dimM$:
\begin{align}
S^{(k)} 
& = \int \dfR^{a_1 a_2}\wedge...\wedge\dfR^{a_{2k-1} a_{2k}}\wedge \star (\cofr_{a_1}\wedge...\wedge \cofr_{a_{2k}}) \nonumber\\ 
& =\frac{(2k)!}{2^k}\int R_{\nu_1\nu_2}{}^{\mu_1 \mu_2}...R_{\nu_{2k-1}\nu_{2k}}{}^{\mu_{2k-1}\mu_{2k}}\nonumber\\
& \qquad\qquad\qquad \delta_{\mu_1}^{[\nu_1}...\delta_{\mu_{2k}}^{\nu_{2k}]} \sqrt{|g|}\mathrm{d}^\dimM x \,, \label{eq: Lovelock general}
\end{align}
where Greek indices refer to a coordinate basis, and Latin indices to a frame in which the metric is the Minkowski metric ($g_{ab}=\eta_{ab}$). As noted in \cite{Gurses2020}, for the Gauss-Bonnet term ($k=2$), if we analyze the problem in differential form notation, when varying the action with respect to the coframe $\cofr^a$, we find
\begin{equation}\label{var1}
\star\frac{\delta S^{(k)}}{\delta \cofr^a} = (\dimM-2k)(\dimM-2k-1)!\ J^{(k)}_{ac}\ \cofr^c\,,
\end{equation}
where $J^{(k)}_{ac}$ is a regular tensor built from combinations of the Riemann tensor that differ for each $k$. The second factor comes from the contraction of two Levi-Civita symbols. Therefore, it is of combinatorial nature: it essentially concerns the counting of the number of possible antisymmetric permutations of a collection of indices. Note that the counting process is not a continuous process in which the number of indices being counted (or equivalently the dimension) can take any value; rather, the value must be an integer. Indeed, for \eqref{var1} to be valid, $\dimM$ must be greater than $2k$ because a $(-1)!$ cannot arise from counting possible permutations. Since \eqref{var1} is not valid for $\dimM=2k$, it cannot be stated that the factor $(\dimM-2k)$ is the responsible for the vanishing of \eqref{var1} in $\dimM=2k$. The reason that it vanishes can actually be traced back to the properties of $2k$-forms in $2k$ dimensions \cite{privZanelli}: by writing the Hodge star operator in \eqref{eq: Lovelock general} explicitly, in the critical dimension, we obtain
\begin{equation}
\star (\cofr_{a_1}\wedge...\wedge \cofr_{a_{2k}}) \overset{{\scriptstyle (\dimM=2k)}}{=} F \epsilon_{a_1 ... a_{2k}}\,,
\end{equation}
where $\epsilon_{a_1 ... a_{2k}}$ is the Levi-Civita tensor associated to the Minkowski metric, and $F$ is a non-zero constant that depends on $k$. As a consequence of this and because the curvature factors in the action do not contribute (via spin connection) to the dynamics in Lovelock theories \cite{Mardones1991}, the vielbein equations of motion are identically satisfied. Observe that this is no longer true if $\dimM>2k$ since, in that case, the Hodge dual of $\cofr_{a_1}\wedge...\wedge \cofr_{a_{2k}}$ is not a $0$-form and makes a non-trivial contribution to the equation of motion of the vielbein.

It is also illuminating  to consider \eqref{var1} as a metric variation, {\it i.e.}, to avoid  the differential form notation and work directly with the metric components. Then, the variation
with respect to the metric of a general $k$-th order
Lovelock term in an arbitrary dimension $\dimM\geq 2k$ is not proportional to $(\dimM-4)$, but rather has the form
\begin{equation}
\frac{1}{\sqrt{|g|}}\frac{\delta S^{(k)}}{\delta g^{\mu\nu}} = (\dimM-2k) A_{\mu\nu} + W_{\mu\nu}, \label{eq: AW decomposition}
\end{equation}
where no $\dimM-2k$ factor can be extracted from $W_{\mu\nu}$. For instance, the first-order Lovelock term (the Einstein-Hilbert action) leads to $A^\text{EH}_{\mu\nu}=0$ and $W^\text{EH}_{\mu\nu}=G_{\mu\nu}$, which vanishes in $\dimM=2$. Analogously, decomposing the Riemann tensor into its irreducible pieces (see {\it e.g.} \cite{Ortin2004}), the second-order Lovelock term, {\it i.e.}, the Gauss-Bonnet term, leads to\footnote{The calculations have been checked with xAct \cite{xact} and we leave the notebook as supplementary material for anyone to check it.\label{xactfootnote}}
\begin{align}
A^\text{GB}_{\mu\nu}  = & \frac{\dimM-3}{(\dimM-2)^2} \Big[\frac{2\dimM}{\dimM-1}R_{\mu\nu} R -\frac{4(\dimM-2)}{\dimM-3}R^{\rho\lambda}C_{\mu\rho\nu\lambda}\nonumber\\
& -4R_{\mu}{}^\rho R_{\nu\rho}+2g_{\mu\nu}R_{\rho\lambda}R^{\rho\lambda}-\frac{\dimM+2}{2(\dimM-1)}g_{\mu\nu} R^2 \Big] \,,\label{eq: Amn GaussBonnet}  \\
W^\text{GB}_{\mu\nu} = & 2 \left[ C_{\mu}{}^{\rho\lambda\sigma}C_{\nu\rho\lambda\sigma}-\frac{1}{4} g_{\mu\nu} C_{\tau\rho\lambda\sigma}C^{\tau\rho\lambda\sigma}\right] \,,
\end{align}
where we have introduced the Weyl tensor $C_{\mu\nu\rho\lambda}$. We note that a similar decomposition for the variation of the GB term (characterized by $k=2$) has also been performed in \cite{Gurses2020}. 

Taking this into account, the field equations given by \eqref{action} in arbitrary dimension are\footnote{Since the trace of $W^\text{GB}_{\mu\nu}$ is proportional to $(\dimM-4)$, the divergence disappears from the trace of the equation of motion, although this factorization cannot be made in the full equation.}
\begin{equation}\label{fieldeqs}
G_{\mu\nu}+\frac{1}{\mP^2}\CosmC g_{\mu\nu}+\frac{2\alpha}{\mP^2}\left(A^\text{GB}_{\mu\nu}+\frac{W^\text{GB}_{\mu\nu}}{\dimM-4}\right)=0\,.
\end{equation}
Hence, with the \textit{regularization} made in \cite{Glavan2020}, {\it i.e.}, by evaluating $\dimM=4$ after calculating the equations of motion in arbitrary $\dimM$, the $A^\text{GB}_{\mu\nu}$ term indeed provides a finite non-trivial correction to the Einstein field equations if the coupling constant of the GB term is $\alpha/(\dimM-4)$. However, in this case, the $W^\text{GB}_{\mu\nu}$ term is ill-defined since in general, $W^\text{GB}_{\mu\nu}$ does not go to zero as $(\dimM-4)$. Indeed, the reason $W^\text{GB}_{\mu\nu}$ vanishes in $\dimM=4$ is that the Riemann tensor loses independent components as one lowers the dimension, and in $\dimM=4$, this loss of components implies that $W^\text{GB}_{\mu\nu}$ necessarily vanishes for algebraic reasons, which is analogous to what happens to the Einstein tensor in $\dimM=2$. In other words, the reason these expressions are zero in certain dimensions is that they are \textit{algebraic} identities fulfilled by the curvatures of all metrics in the critical dimension, as opposed to \textit{functional} identities at which one could arrive by a continuous limiting process given a suitable topology. A somewhat simpler example illustrating a variation that vanishes for algebraic reasons is provided by Galileon or interacting massive vector field theories. There, it can be seen that due to the Cayley-Hamilton theorem, the interaction Lagrangian of a given order $k$ identically vanishes for dimensions higher than the critical dimension associated to $k$ \cite{Beltran2016c}. 

We also mention that the authors of \cite{Glavan2020} appeal to an analogy between their method and the method of dimensional regularization commonly employed in quantum field theory. The dimensional regularization method allows to extract the divergent and finite contributions from integrals that are divergent in $\dimM=4$ but non-divergent for higher $\dimM$. The method considers the analytic continuation of such integrals to the complex plane as a function of the complexified dimension $\dimM$ and then takes the limit $\dimM\to 4$ in a manner that allows the divergent and finite contributions of the integrals to be separated. A key aspect that ensures the well-definiteness of dimensional regularization as an analytic continuation is that the regularized integrals are scalar functions\footnote{Typically the tensorial structures within the integrals are extracted from them by employing Lorentz-covariance arguments, and therefore the integral to regularise is always a scalar function.} that have no algebraic structure sensitive to the number of dimensions of the space they are defined in. Note however that this is not the case for the Gauss-Bonnet term, which has a non-trivial tensorial structure that is not well defined for non-integer dimensions. Thus, although the process of dimensional regularization can be defined by using the smooth $\dimM\to4$ limit of the appropriate analytic continuation of the scalar integrals, this fails to be a continuous limiting process when the quantities involved have a non-trivial algebraic structure, such as tensors or $p$-forms do. 

Regarding this issue, it would also be interesting to attempt to find a precise mathematical meaning to the {\it limiting} procedure in the presence of tensor fields only satisfy certain algebraic identities in a particular number of dimensions. This could be done, for instance, by introducing a formal limit (see, {\it e.g.}, \cite{Mazur2001}) and studying its properties, which would, however, be a highly non-trivial task which lies beyond the scope of this work.

\section{Perturbations around maximally symmetric backgrounds}\label{sec:maxsim}

Despite the above considerations, we acknowledge that even though the {\it regularization} method proposed in \cite{Glavan2020} does not work in general, it suffices for finding solutions that satisfy enough symmetries so as to render the $W^\text{GB}_{\mu\nu}$ identically zero in an arbitrary dimension. Thus, by symmetry-reducing the action before enforcing $\dimM=4$, we eliminate the problematic $W^\text{GB}_{\mu\nu}$ term and arrive at well-defined equations of motion. This is the case, for instance, of all conformally flat geometries, which have an identically vanishing Weyl tensor in $D\geq4$, thus satisfying the desired property that $W^\text{GB}_{\mu\nu}=0$ in $\dimM\geq4$, which makes the $D\to4$ {\it limit} of the (symmetry-reduced) D4EGB field equations \eqref{fieldeqs} well defined. Maximally symmetric geometries or FLRW spacetimes are conformally flat, and therefore, the \textit{regularized} D4EGB equations are well defined in such situations. Let us analyze the maximally symmetric solutions of \eqref{fieldeqs} studied in \cite{Glavan2020}. In these geometries, the Riemann tensor is given by
\begin{equation}\label{eq: max symm}
R_{\mu\nu}{}^{\rho\sigma}=\frac{\KMaxSym}{\mP^2(\dimM-1)}\left(\delta^{\rho}_{\mu} \delta^{\sigma}_{\nu}-\delta^{\sigma}_{\mu} \delta^{\rho}_{\nu}\right),
\end{equation}
and $W^\text{GB}_{\mu\nu}$ vanishes in an arbitrary dimension as explained above. In this case, the variation of the GB term is indeed proportional to $(\dimM-4)$, and therefore, after the symmetry-reduction of the action \eqref{action}, the field equations \eqref{fieldeqs} read
\begin{equation}\label{fieldeqssym}
   G_{\mu\nu}+\frac{1}{\mP^2}\CosmC g_{\mu\nu}+\frac{2\alpha}{\mP^2}A^\text{GB}_{\mu\nu}=0 \,,
\end{equation}\label{regularizedEOM}
where $A^\text{GB}_{\mu\nu}$ provides a regular  $\alpha-$dependent correction to GR. Although these properties are shared by any conformally flat solution, one should bear in mind that arbitrary perturbations around these backgrounds are sensitive to the ill-defined contributions that come from the $W^\text{GB}_{\mu\nu}$ dependence of the full D4EGB field equations \eqref{fieldeqs}.

It is worth noticing, though, that the ill-defined corrections that enter the equations of motion through the $\alpha W^\text{GB}_{\mu\nu}/(\dimM-4)$ term do not contribute to linear order in perturbation theory around a maximally symmetric background. Presumably, this is the reason why these problematic contributions went unnoticed in \cite{Glavan2020}, where only linear perturbations were considered. Nonetheless, the ill-defined terms related to $W^\text{GB}_{\mu\nu}$ will enter the perturbations at second-order.
 
To show this, let us consider a general perturbation around a maximally symmetric background by splitting the full metric as
\begin{equation}
    g_{\mu\nu}=\bar{g}_{\mu\nu}+\epsilon h_{\mu\nu}
\end{equation}
where $\bar{g}_{\mu\nu}$ is a maximally symmetric solution of \eqref{fieldeqs}. Therefore, the left hand side of \eqref{fieldeqs} can be written as a perturbative series in $\epsilon$ of the form
\begin{equation}
     E^{(0)}{}_{\mu\nu}+\epsilon E^{(1)}{}_{\mu\nu}+\epsilon^2 E^{(2)}{}_{\mu\nu}\ldots\,,
\end{equation}
where $E^{(0)}{}_{\mu\nu}=0$ are the background field equations, $E^{(1)}{}_{\mu\nu}=0$ are the equations for linear perturbations, and so on. Using the zeroth-order equation, the linear perturbations in $\dimM$ dimensions and around a maximally symmetric background are described by\footnote{Although \eqref{linearpert} and the equations for linear perturbations in \cite{Glavan2020} differ by the ordering of the covariant derivatives of the $\nabla_{\rho}\nabla_{\nu}h^{\mu\rho}$ term and the sign in the mass term, our equation \eqref{fieldeqs} coincides with those in {\it e.g.} \cite{Ortin2004} for linearized perturbations around a maximally symmetric background.}
\begin{align}
0=&\left(1+\frac{4(\dimM-3)}{\dimM-1}\frac{\alpha\KMaxSym}{\mP^4}\right)\times \nonumber\\
&\bigg[\nabla^{\rho}\nabla^{\mu}h_{\nu\rho}+\nabla_{\rho}\nabla_{\nu}h^{\mu\rho}-\nabla^{\rho}\nabla_{\rho} h_{\mu\nu}-\nabla^{\mu}\nabla_{\nu}h \nonumber\\
& +\delta^{\mu}{}_{\nu}(\nabla^{\sigma}\nabla_{\sigma}h -\nabla_{\rho}\nabla_{\sigma}h^{\rho\sigma})-\frac{\KMaxSym}{\mP^2} (\delta^\mu_\nu h-2h^\mu{}_\nu)\bigg]\,,\label{linearpert}
\end{align}
where $h\equiv h^\sigma{}_\sigma$ and the indices have been raised by $\bar g^{\mu\nu}$. By inspection, we can see that this equation is regular in $\dimM=4$. Furthermore, as noted in \cite{Glavan2020}, the equation governing linear perturbations \eqref{linearpert} is essentially that of GR, although multiplied by an overall factor that depends on $\alpha$. Let us now consider the quadratic order in the perturbations. For our purpose, it will be sufficient to consider quadratic perturbations around a Minkowskian background. By using the zeroth- and first-order perturbation equations, and enforcing a vanishing background curvature $\KMaxSym=0$, we can write the second-order perturbation equations $E^{(2)}{}_{\mu\nu}=0$ as (see supplementary material)
\begin{align}\label{quadraticpert}
0 & =[\text{GR terms of }\mathcal{O}(h^{2})]_{\mu\nu} +\frac{\alpha}{\mP^{2}(\dimM-4)} \times \nonumber \\
 & \quad\Big\{
 -2\nabla_{\gamma}\nabla_{\alpha}h_{\nu\beta}\nabla^{\gamma}\nabla^{\beta}h_{\mu}{}^{\alpha}
 +2\nabla_{\gamma}\nabla_{\beta}h_{\nu\alpha}\nabla^{\gamma}\nabla^{\beta}h_{\mu}{}^{\alpha}\nonumber \\
 &\qquad +2\nabla^{\gamma}\nabla^{\beta}h_{\nu}{}^{\alpha}\nabla_{\mu}\nabla_{\alpha}h_{\beta\gamma} 
 +2\nabla^{\gamma}\nabla^{\beta}h_{\mu}{}^{\alpha}\nabla_{\nu}\nabla_{\alpha}h_{\beta\gamma}\nonumber \\
 &\qquad-2\nabla^{\gamma}\nabla^{\beta}h_{\mu}{}^{\alpha}\nabla_{\nu}\nabla_{\beta}h_{\alpha\gamma} 
 -2\nabla^{\gamma}\nabla^{\beta}h_{\nu}{}^{\alpha}\nabla_{\mu}\nabla_{\beta}h_{\alpha\gamma}\nonumber \\
 &\qquad
 -2\nabla_{\mu}\nabla^{\gamma}h^{\alpha\beta}\nabla_{\nu}\nabla_{\beta}h_{\alpha\gamma}
 +2\nabla_{\mu}\nabla^{\gamma}h^{\alpha\beta}\nabla_{\nu}\nabla_{\gamma}h_{\alpha\beta}\nonumber \\
 & \qquad +g_{\mu\nu}\big(
 2\nabla_{\delta}\nabla_{\beta}h_{\alpha\gamma}\nabla^{\delta}\nabla^{\gamma}h^{\alpha\beta}
 -\nabla_{\delta}\nabla_{\gamma}h_{\alpha\beta}\nabla^{\delta}\nabla^{\gamma}h^{\alpha\beta}\nonumber \\
 & \qquad\qquad -\nabla_{\beta}\nabla_{\alpha}h_{\gamma\delta}\nabla^{\delta}\nabla^{\gamma}h^{\alpha\beta}\big) \Big\}\,.
\end{align}
Notice that, given that the numerator of the $1/(D-4)$ term comes entirely from the $W^\text{GB}_{\mu\nu}$ contribution in \eqref{eq: AW decomposition}, it vanishes identically in $\dimM=4$, rendering an indeterminate $0/0$ in the second order perturbation equations after the {\it limit} outlined by Glavan and Lin in \cite{Glavan2020} is taken. 

In essence, the problem with the {\it limiting} prescription outlined in \cite{Glavan2020} is that it does not make sense of a general tensor as a continuous function of the dimension $D$ of the space in which it is defined. Evidently, if a suitable prescription to treat the numerator as an analytic function of $\dimM$ was found and went as $(\dimM-4)^n$ with $n\geq 1$, then the {\it limit} would be well-defined. However, this would require to give a precise definition for the limit of a tensor as a function of the spacetime dimension, which is not  given in \cite{Glavan2020}. 

In other words, the pathological term in the full equation of motion, which does not appear at linear order around Minkowski spacetime, enters at the next  order in the perturbative expansion, providing a clear example that suggests that the D4EGB field equations \eqref{fieldeqs} are generally ill-defined. This results are somewhat in the line of that found in \cite{Bonifacio2020}, where it was seen that the amplitudes of GB in the $\dimM\to4$ \textit{limit} correspond to those of a scalar-tensor theory, in which the scalar is infinitely strongly coupled (which suggests that the pathological term in the second order perturbations actually blows up). Hence, they concluded that this new pathological degree of freedom would only appear beyond linear order perturbations.

Moreover, going beyond a Minkowskian background, perturbations around arbitrary maximally symmetric backgrounds \eqref{eq: max symm} pick up additional $\Lambda$-proportional terms which contain a factor $1/(\dimM-4)$ (see the supplementary material). Concretely, up to second order in $h_{\mu\nu}$, there are $\mathcal{O}(\Lambda)$ terms of the form $h(\nabla^2 h)$ and $\mathcal{O}(\Lambda^2)$ terms of the form $h^2$. Consequently, the $\Lambda$-proportional terms provide additional undetermined terms that make (anti-)de Sitter backgrounds pathological.

\section{An action for the regularized equations?}\label{sec:action}

We have seen that unless the field equations \eqref{fieldeqs} are stripped of the $W^\text{GB}_{\mu\nu}$ term after taking the variation of the D4EGB action \eqref{action}, they are, in general, ill-defined. Let us now comment on the possibility of finding a diffeomorphism-invariant action whose field equations in $\dimM\geq4$ are of the form \eqref{fieldeqssym}.\footnote{Even though the $\dimM\to4$ process, if understood as a limit, will have the same conceptual problems described in section \ref{sec:Dto4}, in this case they might be swept under the rug since the $1/(\dimM-4)$ dependence actually disappears from the field equations.}

To find such an action starting from the EGB one, we should be able to subtract a scalar from the EGB action so that the contribution of $W^\text{GB}_{\mu\nu}$ disappears after taking the variation with respect to the metric, without losing the diffeomorphism symmetry of the EGB action. In trying to find such a term, we immediately arrive at an inconsistency, which we proceed to illustrate. Diffeomorphism invariance of the Gauss-Bonnet action, {\it i.e.} \eqref{eq: Lovelock general} with $k=2$, implies that its variation with respect to the metric is divergenceless.\footnote{This is due to the Bianchi identity under diffeomorphisms, see {\it e.g.} \cite{Ortin2004}.} Thus, by using the A-W decomposition \eqref{eq: AW decomposition} and substituting $A^\text{GB}_{\mu\nu}$ with \eqref{eq: Amn GaussBonnet}, the off-shell relation follows
\begin{align}\label{DivW}
      \nabla^\mu W^\text{GB}_{\mu\nu} = - \frac{4(\dimM-4)}{\dimM-2}C_{\nu\rho\lambda\mu}\nabla^\mu R^{\rho\lambda}\,.
\end{align}
Observe that the right-hand side of this equation is not identically zero in an arbitrary dimension, as can be seen by considering the following counterexample in five dimensions:
\begin{equation}
    {\rm d}s^2={\rm d}t^2 - {\rm e}^{2t}{\rm d}x^2- {\rm e}^{4t}({\rm d}y^2+{\rm d}z^2+{\rm d}w^2)\,,
\end{equation}
for which equation \eqref{DivW} reads
\begin{equation}
    \nabla^\mu W^\text{GB}_{\mu\nu} = -4 \delta^t_\nu\neq0\,.
\end{equation}

Together with the fact that the variation with respect to the metric of any diffeomorphism-invariant action is identically divergence-free, the above result implies that the $W^{\text{GB}}_{\mu\nu}$ term does not come from an action that is a scalar under diffeomorphisms. Consequently, there does not exist any term that can be added to the action \eqref{action} to cancel the $W^{\text{GB}}_{\mu\nu}$ contribution in the D4EGB field equations \eqref{fieldeqs} without losing its diffeomorphism-invariance. 
Other authors have proposed alternative ways to regularize the action \eqref{action}, generally leading to a scalar-tensor theory  of  the  Horndeski  family \cite{Lu2020,Bonifacio2020,Fernandes2020,Hennigar2020},  thus  leaving  the Lovelock theorem intact.

We thus conclude that no diffeomorphism-invariant action can give the desired field equations \eqref{fieldeqssym} in $\dimM\geq4$. Nevertheless, nothing precludes the existence of a non-diffeomorphism-invariant action having \eqref{fieldeqssym} as its field equations. Where it possible to find such an action, however, the absence of diffeomorphism invariance would potentially unleash the well-known pathologies that occur in massive gravity (see, {\it e.g.,} \cite{Hinterbichler2012,deRham2014}), thus propagating a Boulware-Deser ghost \cite{Boulware1972}. 

\section{Geodesic analysis of the spherically symmetric solutions}\label{sec:geodesic}

In addition to maximally symmetric and FLRW spacetimes, spherically symmetric solutions of D4EGB are also considered in \cite{Glavan2020}, where it is stated that they are described by the 4-dimensional metric 
\begin{equation}\label{sphericallysymmetricmetric}
    {\rm d}s^2 = A_{\pm}(r){\rm d}t^2 -A_{\pm}^{-1}(r){\rm d}r^2-r^2{\rm d}\Omega^2_2\,,
\end{equation}
where $A_{\pm}(r)$ has the form
\begin{equation}\label{eq:metric}
A_{\pm}(r)=1+\frac{r^2}{32\pi\alpha G}\left[1\pm\sqrt{1+\frac{128\pi\alpha G^2 M}{r^3}}\right].
\end{equation}
First, we note that $\dimM-$dimensional spherically symmetric geometries described by metrics of the form \cite{Ortin2004}
\begin{equation}\label{sphericalmetricD}
{\rm d}s^2 = A(r){\rm d}t^2 -A^{-1}(r){\rm d}r^2-r^2{\rm d}\Omega^2_{\dimM-2}\,,
\end{equation}
do not in general satisfy $W^{\text{GB}}_{\mu\nu}=0$ in arbitrary $\dimM\geq4$. To see this, it suffices to restrict consideration to the 5-dimensional case, where the condition for $W^{\text{GB}}_{\mu\nu}$ to vanish is
\begin{equation}
    r^2\frac{{\rm d}^2A}{{\rm d}r^2}-2r\frac{{\rm d}A}{{\rm d}r}+2A-2=0\,.
\end{equation}
This only happens for the particular case \mbox{$A=1+C_1 r+C_2 r^2$}, where $C_i$ are integration constants. This suggests that \eqref{eq:metric} cannot be regarded as a solution of the D4EGB field equations, given that \eqref{fieldeqs} is not well-defined for $\dimM$-dimensional spherically symmetric metrics \eqref{sphericalmetricD} in the $\dimM\to4$ \textit{limit}. Indeed, as the authors of \cite{Glavan2020} explain, the 4-dimensional spherically symmetric geometries  \eqref{eq:metric} are obtained by first re-scaling $\alpha$ by a factor of $1/(\dimM-4)$ in the solutions obtained in \cite{Boulware1985} for EGB in $\dimM\geq5$ and then taking the $\dimM\to4$ \textit{limit}, instead of solving the $\dimM\to4$ \textit{limit} of \eqref{fieldeqs}.

Nevertheless, it could be that the spherically symmetric geometries of \cite{Glavan2020} are solutions of \eqref{fieldeqssym}, that is, of the field equations \eqref{fieldeqs} after being stripped of the pathological $W^{\text{GB}}_{\mu\nu}$ term. In the supplementary material, it can be seen that \eqref{fieldeqssym} has four different branches of solutions for $\alpha>0$. Two of them are exactly the Schwarzschild and Schwarzschild-(anti-)de Sitter solutions:
\begin{align}
    &A_1=1-\frac{2GM}{r}\,,\nonumber\\
    &A_2=1+\frac{r^{2}}{16\pi G \alpha}-\frac{2GM}{r}\,.
\end{align}
and the other two cannot be solved analytically, though their asymptotic behavior near the origin can be seen to be $A\cong r^{-3-2\sqrt{3}}+\mathcal{O}(r^0)$. Therefore, these solutions can not be the ones found in \cite{Glavan2020} either, although they approach the Schwarzschild and Schwarzschild-(anti-)de Sitter solutions at spatial infinity.

Let us now turn to the behavior of the spherically symmetric geometries presented in \cite{Glavan2020}. As noted in \cite{Glavan2020}, the $\alpha<0$ branch of the above solution is not well defined {for values of the radial coordinate below $r<(-128\pi\alpha G^2 M)^{-1/3}$}, so their analysis focuses on the $\alpha>0$ branches, showing that the above metric describes solutions that behave asymptotically as Schwarzschild or Schwarzschild-de Sitter solutions by choosing negative and positive signs, respectively.

Concerning the former branch of solutions, it is shown in \cite{Glavan2020} that {its} causal structure {(namely, the presence or absence of event horizons)} depends on the ratio between the {\it mass} {parameter} $M$ and a new mass scale $M_*=\sqrt{16\pi\alpha/G}$, which characterizes the D4EGB corrections to GR. From \eqref{sphericallysymmetricmetric} and \eqref{eq:metric}, it can be shown that the $g_{tt}$ component of the metric {vanishes at the spherical surfaces}:
\begin{equation}\label{eq:horizons}
    r_\pm=GM\left[1\pm\sqrt{1-\left(\frac{M_*}{M}\right)^2}\right].
\end{equation}
In view of this expression, it becomes clear that solutions have no horizons if $M<M_*$, outer and inner horizons if $M>M_*$, and one degenerate horizon if $M=M_*$. Interestingly, the mass scale $M_{*}$ plays a role similar to that of the electric (and magnetic) charges in the Reissner-Nordstr{\"o}m spacetime, with the exception that in this case, the origin of such contributions comes exclusively from the gravitational field. The effect of the Gauss-Bonnet terms is that of making gravity repulsive at short distances, the magnitude of this repulsion being dictated by the strength of the GB coupling $\alpha$.

{Regarding the presence of singularities in the solutions, we see that despite the metric components \eqref{eq:metric} being finite at the origin}: \begin{equation}
A(r)=1-\sqrt{\frac{2M}{GM_*^2}}r^{1/2}+\mathcal{O}(r^{3/2})\,,  
\end{equation}
the curvature invariants diverge as $R\propto r^{-3/2}$, $R_{\mu\nu}R^{\mu\nu}\sim R_{\mu\nu\alpha\beta}R^{\mu\nu\alpha\beta}\propto r^{-3}$. {In \cite{Glavan2020} it is argued that an observer could never reach this curvature singularity given the repulsive effect of gravity at short distances.} {This would imply} that the spacetime described by \eqref{sphericallysymmetricmetric} is complete in the sense that no (classical) physical observer ever reaches the curvature singularity at $r=0$ in a finite proper time. Nonetheless, there is no explicit proof in \cite{Glavan2020} showing that this is indeed the case. We thus proceed to answer precisely the following question: does any (classical) physical observer reach the curvature singularity of \eqref{sphericallysymmetricmetric} in a finite proper time? To answer this question, it suffices to study the sub-class of radial freely-falling (classical) observers, described by time-like geodesics. We also consider radial null geodesics for completeness. 

\begin{figure}
    \centering
    \includegraphics[width=\columnwidth]{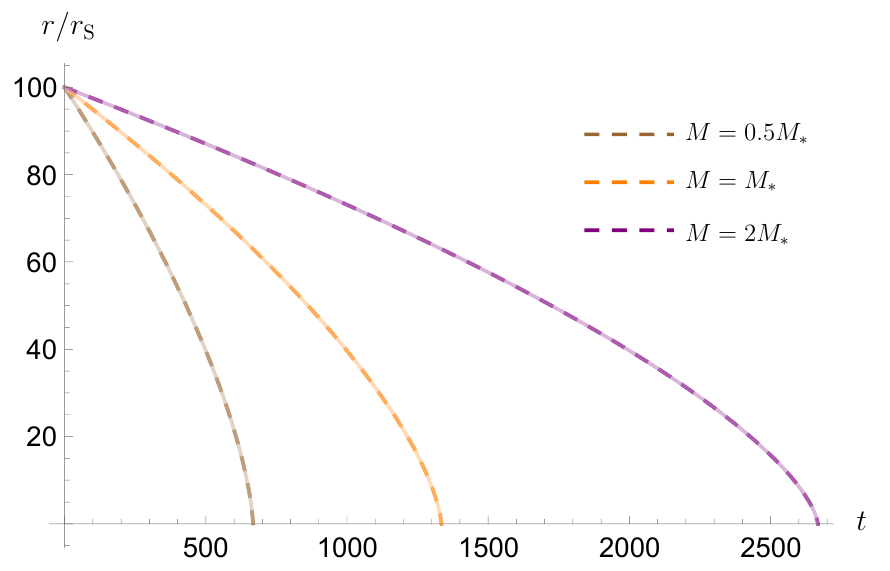}
    \caption{Plot of the radial ingoing trajectories (in units of $r_{\rm S}=2M$) of a free-falling massive particle in the spacetime \eqref{sphericallysymmetricmetric} (dashed lines) and of the Schwarzschild solution (pastel colors) for different values of the parameter $M$. At large distances, the trajectories are indistinguishable. We use $M_{*}=G=1$, $r(0)=100~r_{\rm S}$, and $E=1$ for visualization purposes.}
    \label{IngoingGeodesics}
\end{figure}

\begin{figure}
    \centering
    \includegraphics[width=\columnwidth]{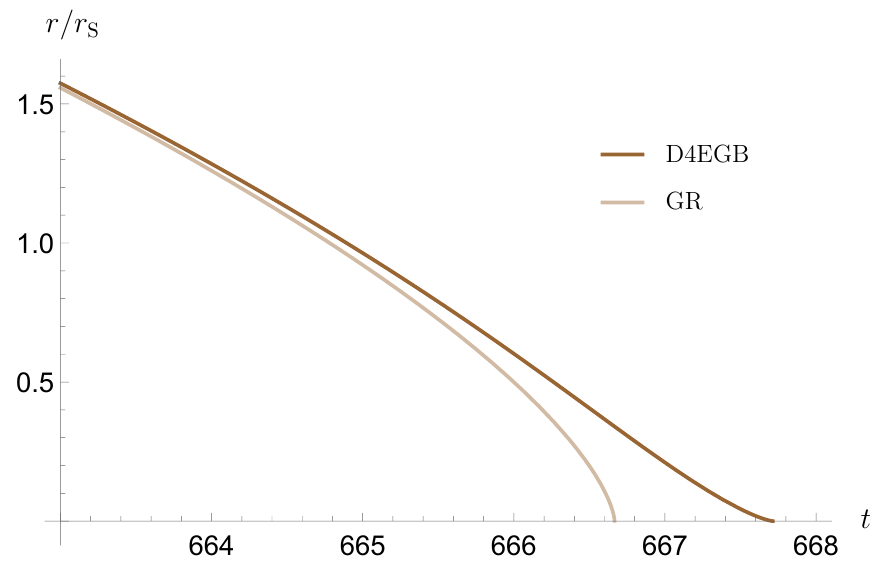}
    \caption{Figure showing how the trajectories of a massive particle in the spacetime described by \eqref{sphericallysymmetricmetric} (in units of \mbox{$r_{\rm S}=2M$}) deviate from those of the Schwarzschild solution from GR near the central curvature singularity. Here, only the case $M=0.5M_*$ is shown, although all timelike geodesics exhibit the same behavior near the singularity. The units are $M_{*}=G=1$.}
    \label{IngoingGeodesicsnear0}
\end{figure}

 Consider the geodesic equation in the equatorial plane\footnote{Since spacetime is spherically symmetric, geodesics will lie in a plane, which can be chosen as the equatorial one in suitable coordinates. See {\it e.g.} \cite{Wald1984} for details on the derivation of the geodesic equation and \cite{Olmo2015} for the completeness analysis. } $\theta=\pi/2$ for the metric \eqref{sphericallysymmetricmetric}
\begin{equation}\label{geoeq}
\left(\frac{{\rm d}r}{{\rm d}\tau}\right)^2=E^2-V_{\text{eff}}(r), 
\end{equation}
with
\begin{equation}
V_{\text{eff}}(r)=A(r)\left(\frac{L^2}{r^2}-\kappa\right),
\end{equation}
where $\tau$ is the proper time of the observer that moves along the solution $r(\tau)$. Here, $\kappa$ takes the values $\{-1,1,0\}$ for space-like, time-like, and null geodesics, respectively. $E$ and $L$ are constants of motion associated with time-translation and rotational symmetries, respectively. For our purpose, it suffices to analyze radial geodesics, characterized by $L=0$. First note that since photon trajectories are {insensitive} to the value of $A(r)$ in a spacetime described by any metric of the form \eqref{sphericallysymmetricmetric}, the trajectories stay the same as in GR. The solution to \eqref{geoeq} for time-like geodesics is plotted in fig. \ref{IngoingGeodesics} for cases with different causal structures. There, it can be seen that infalling massive particles starting in a region well beyond the Schwarzschild radius (where the space-time is effectively the same as in GR) reach the curvature singularity at $r=0$ in a finite proper time (no matter the initial velocity). Notice that, as can be seen in fig. \ref{IngoingGeodesicsnear0}, the deviations that form the GR trajectories are not relevant until the particle is at $r\cong r_{S}$. An asymptotic analysis of the geodesic equation reveals that in GR, the curvature singularity at $r=0$ is reached with infinite velocity ${\rm d}r/{\rm d}\tau|_{\text{GR}}\propto r^{-1/2}+\mathcal{O}(r^{0})$, while the geodesics described by \eqref{sphericallysymmetricmetric} reach it with finite velocity:
\begin{equation}\label{eq:velocity}
  \left.\left(\frac{{\rm d}r}{{\rm d}\tau}\right)^2\right|_{\text{D4EGB}}=E^{2}-1+\sqrt{\frac{2M}{M_*^{2}}}r^{1/2}+\mathcal{O}(r^{3/2})\,.
\end{equation}
It is interesting to note that if the infalling particle starts at rest, no matter what its initial position is, it will reach the singularity with zero velocity (characterized by $E^2=1$): attractive and repulsive effects compensate for each other along the trajectory of the particle.
The above proves that the statement made in \cite{Glavan2020} that particles cannot reach the central singularity in spacetimes described by \eqref{sphericallysymmetricmetric} is not correct, as the singularity is reached in finite affine parameter.  Therefore, the hope that these solutions avoid the singularity problem is cast into serious doubt. The authors of \cite{Glavan2020} also claim that under a realistic stellar collapse, matter would stop before reaching the singularity. This must be verified by a self-consistent analysis of the dynamical collapsing geometry, as reported in \cite{Malafarina2020}, revealing that the singularity indeed forms and gets covered by a horizon. Furthermore, the authors of \cite{Malafarina2020} found that if the collapse is modelled {\it \`a la} Oppenheimer-Snyder, where the dust is initially at rest, matter reaches the singularity with zero velocity, in agreement with our results.

We also note that even if geodesic observers never arrive at the singularity, the usual problems regarding curvature singularities would remain: quantum corrections would be expected to become non-perturbative near the singularity, and the background could not be treated classically anymore. However, the solutions would be {\it classically} singularity free in this case.

\section*{Final remarks} 

We have investigated the idea of providing corrections to four dimensional General Relativity by means of the Gauss-Bonnet term analyzed and devised in \cite{Tomozawa2011} and recently revisited in \cite{Glavan2020}. We have shown that this idea cannot be implemented for the Gauss-Bonnet ($k$-th order Lovelock) term in four ($2k$) spacetime dimensions by means of the procedure considered in \cite{Glavan2020} without encountering inconsistencies.

When considering solutions with a high degree of symmetry, such as maximally symmetric or general conformally flat solutions, this issue is concealed at the level of the equations of motion because the problematic terms $W^\text{GB}_{\mu\nu}$ in \eqref{fieldeqs} vanish for arbitrary $\dimM$ in these scenarios. Indeed, we have shown that when considering perturbations around a Minkowskian (or any maximally symmetric) background beyond linear order, such inconsistencies are immediately unveiled. This also aligns with the conclusions that the authors of \cite{Bonifacio2020} arrived at by analysing the GB amplitudes. 

Regarding the spherically symmetric geometries presented in \cite{Glavan2020}, we showed that they do not attain the required degree of symmetry as to make $W^\text{GB}_{\mu\nu}$ vanish in arbitrary dimension and thus bypass the pathologies encountered in the field equations \eqref{fieldeqs}. By artificially removing $W^\text{GB}_{\mu\nu}$ from \eqref{fieldeqs}, we encountered four spherically symmetric solutions, none of which coincides with those presented in \cite{Glavan2020}. Moreover, a geodesic analysis of the geometries from \cite{Glavan2020} contradicts the observation about the singularity being unreachable by any observer in finite proper time.

The idea of extracting non-trivial corrections to the dynamics of a theory from topological terms by considering a divergent coupling constant is indeed very appealing since its range of applicability extends far beyond gravitational contexts. For instance, it might serve to introduce parity-violating effects in Yang-Mills theories through the corresponding $F\tilde F$ terms that are topological in four dimensions. Indeed, a similar idea has been seen to lead to well-defined theories in the context of Weyl geometry \cite{Beltran2014,Beltran2016a,Beltran2016b}. It could thus be interesting to explore various possibilities in this direction.

\section*{Acknowledgements}
The authors are very grateful to Carlos Barcel\'o, Jose Beltr\'an Jim\'enez, Bert Janssen, Gonzalo J. Olmo, Jorge Zanelli, and an anonymous referee for their useful comments. AJC and AD are supported by PhD contracts of the program FPU 2015 with references FPU15/02864 and FPU15/05406 (Spanish Ministry of Economy and Competitiveness), respectively. This work is supported by the Spanish Projects No. FIS2017-84440-C2-1-P (MINECO/FEDER, EU) and FIS2016-78198-P (MINECO), the Project No. H2020-MSCA-RISE-2017 GrantNo. FunFiCO-777740, Project No. SEJI/2017/042 (Generalitat Valenciana), the Consolider Program CPANPHY-1205388, and the Severo Ochoa Grant No. SEV-2014-0398 (Spain). JA acknowledges financial support from the Spanish Government through the project FIS2017-86497-C2-2-P (with FEDER contribution), and from the State Agency for Research of the Spanish MCIU through the ``Center of Excellence Severo Ochoa'' award to the Instituto de Astrof\'{\i}sica de Andaluc\'{\i}a (SEV-2017-0709).

\end{document}